%% file: UAV_WCM.tex
\documentclass[final]{IEEEtran}
\newlength \figwidth
\usepackage{cite}
\ifCLASSINFOpdf
  \usepackage[pdftex]{graphicx}
	\usepackage{epstopdf}
  \graphicspath{{../Figures/}}
  \DeclareGraphicsExtensions{.eps}
\else
  \usepackage[dvips,draft]{graphicx}
\fi
\usepackage[cmex10]{amsmath}
\usepackage[table]{xcolor} 
\usepackage{amssymb}
\usepackage{pifont}
\usepackage{amsthm}
\usepackage{url}
\usepackage{dsfont}
\usepackage{tabulary}
\usepackage{multirow}
\usepackage{color}

\usepackage{color}
\usepackage{times}
\usepackage{verbatim}
\usepackage{floatflt}
\usepackage{enumerate}
\usepackage{array}
\usepackage{multicol,afterpage,wrapfig} 
\usepackage{tikz}
\usepackage{algorithm}
\usepackage[font=small]{caption}
\usepackage[noend]{algpseudocode}
\makeatletter
\def\BState{\State\hskip-\ALG@thistlm}
\makeatother
\usepackage{amsmath,amssymb,eucal}
\usepackage[utf8]{inputenc}
\usepackage{epsfig}
\usepackage{exscale}
\usepackage{latexsym}
\usepackage{verbatim}
\usepackage{amsfonts}
\usepackage{multirow}
\usepackage{cite}
\usepackage{balance}
\usepackage{color,colortbl}
\usepackage{transparent}
\usepackage{import}
\usepackage{mathtools}
\usepackage{tikz}
\usetikzlibrary{automata,arrows,positioning,calc}
\usepackage{bbm}
\usepackage[printonlyused,withpage]{acronym}
\usepackage{tabu}
\usepackage{longtable}
\usepackage{balance}
\usepackage{footnote}
\usepackage{tablefootnote}
\usepackage{subfig}
\definecolor{Gray}{gray}{0.95}
\definecolor{LightCyan}{rgb}{0.8,0.85,1}
\usepackage{pifont}
\newcommand{\cmark}{\ding{51}}%
\newcommand{\xmark}{\ding{55}}%

\def\BibTeX{{\rm B\kern-.05em{\sc i\kern-.025em b}\kern-.08em
    T\kern-.1667em\lower.7ex\hbox{E}\kern-.125emX}}
    
\renewcommand{\IEEEQED}{\IEEEQEDopen}

\newcommand*\xbar[1]{%
  \hbox{%
    \vbox{%
      \hrule height 0.5pt 
      \kern0.36ex
      \hbox{%
        \kern-0.12em
        \ensuremath{#1}%
        \kern-0.12em
      }%
    }%
  }%
}

\input{macros}

\allowdisplaybreaks

\IEEEoverridecommandlockouts
\begin{document}
\pagenumbering{gobble}

\newtheorem{Theorem}{\bf Theorem}
\newtheorem{Corollary}{\bf Corollary}
\newtheorem{Remark}{\bf Remark}
\newtheorem{Lemma}{\bf Lemma}
\newtheorem{Proposition}{\bf Proposition}
\newtheorem{Assumption}{\bf Assumption}
\newtheorem{Definition}{\bf Definition}
\title{The Essential Guide to Realizing\\5G-Connected UAVs with Massive MIMO}
\author{\IEEEauthorblockN{{Adrian~Garcia-Rodriguez, Giovanni~Geraci, David~L\'{o}pez-P\'{e}rez,\\Lorenzo~Galati~Giordano, Ming Ding, and Emil Bj\"{o}rnson}}
\thanks{A.~Garcia-Rodriguez, G.~Geraci, D.~L\'{o}pez-P\'{e}rez, and L.~Galati~Giordano are with Nokia Bell Labs, Ireland (\{adrian.garcia\_rodriguez, giovanni.geraci, david.lopez-perez, lorenzo.galati\_giordano\}@nokia-bell-labs.com).}
\thanks{M.~Ding is with Data61, CSIRO, Australia (ming.ding@data61.csiro.au).}
\thanks{E.~Bj\"{o}rnson is with the Department of Electrical Engineering (ISY), Link\"{o}ping University, Sweden (emil.bjornson@liu.se).}
}
\maketitle
\thispagestyle{empty}
\input{Abstract}
\IEEEpeerreviewmaketitle
\input{Section1}
\input{Section2}
\input{Section3}
\input{Section4}
\input{Section5}
\input{Section6}
\balance
\ifCLASSOPTIONcaptionsoff
  \newpage
\fi
\bibliographystyle{IEEEtran}
\bibliography{Strings_Gio,Bib_Gio}
\end{document}

%% file: macros.tex
\newfont{\bbb}{msbm10 scaled 500}

\newfont{\bb}{msbm10 scaled 1100}

\newcommand{\executeiffilenewer}[3]{%
\ifnum\pdfstrcmp{\pdffilemoddate{#1}}%
{\pdffilemoddate{#2}}>0%
{\immediate\write18{#3}}\fi%
}

%% file: Abstract.tex
\begin{abstract}
What will it take for drones---and the whole associated ecosystem---to \emph{take off}? Arguably, infallible command and control (C\&C) channels for safe and autonomous flying, and high-throughput links for multi-purpose live video streaming. And indeed, meeting these aspirations may entail a full cellular support, provided through 5G-and-beyond hardware and software upgrades by both mobile operators and manufacturers of these unmanned aerial vehicles (UAVs).
In this article, we vouch for massive MIMO as the key building block to realize 5G-connected UAVs. Through the sheer evidence of 3GPP-compliant simulations, we demonstrate how massive MIMO can be enhanced by complementary network-based and UAV-based solutions, resulting in consistent UAV C\&C support, large UAV uplink data rates, and harmonious coexistence with legacy ground users.
\end{abstract}

%% file: Section1.tex
\section*{Introduction}

Believe it or not, there will likely be a drone for everyone in the years to come. 
If you are a daredevil climber, it may make you feel safer knowing that drones could facilitate search-and-rescue missions, should something go wrong up in the mountains. 
If sport is not your thing, and you would rather sit in front of a TV, a drone may be shooting your favorite documentary or delivering your piping-hot take-away pizza. The advantages of drones performing such, and many more, vital functions are easy to visualize, but \emph{what will it take?} 
One of the key answers lies in wireless communications: 
reliable command and control (C\&C) channels allowing autonomous drone cruising---whether in the woods or in downtown Manhattan---paired with high-data-rate connections enabling real-time streaming of events like political rallies, traffic jams, or the Tour de France \cite{HayYanMuz2016,ChaLar2017,YanLinLi2018}.

Mobile network operators (MNOs) are well aware of, and lured by, the new revenue opportunities stemming from a proliferation of drones---also known by the most tech savvy as unmanned aerial vehicles (UAVs). 
MNOs are thus, more than ever, 
\emph{preparing the ground} to tackle this new vertical market by offering cellular coverage to a heterogeneous population of users, 
comprising both terrestrial and aerial equipment. 
Every operator's aspiration would be to support C\&C and data channels of a large number of UAV users by seamlessly reusing existing, or soon-to-be-deployed, network infrastructure. However, studies undertaken by academia \cite{MozSaaBen2018,ZenLyuZha2018,AzaRosPol2017,BerChiPol2016} and key industry players \cite{LinYajMur2018,YajWanGao2018,GerGarGal2018,NguAmoWig2018} have unanimously pointed out that important technical challenges may have to be overcome for cellular-connected UAVs to be more than just wishful thinking.

In this article, we adopt the view of an MNO that, by rolling out its massive MIMO-based 5G network, aims to offer cellular communication services to both ground users (GUEs) and UAVs simultaneously. With this purpose, our MNO intends to reuse its already-purchased 10~MHz of sub-6~GHz licensed spectrum, operating it in time division duplexing (TDD) mode. 
The MNO under consideration is particularly interested in assessing the performance of its fully loaded network in an Urban Macro scenario, where it has deployed three-sector base stations (BSs), 500~m apart at a height of 25~m, to provide cellular coverage to an average of 15 active devices per sector.

\emph{Here is the dilemma:} 
Will this infrastructure suffice to meet the UAVs' link requirements---100~kbps C\&C channel and uplink payloads demanding several Mbps---set forth by the standardization forum?\footnote{The Third Generation Partnership Project (3GPP) \cite{3GPP36777}.} Or should the MNO's network, primarily catering to GUEs, undergo substantial upgrades?

The only way to provide well founded answers to such questions is by accurately evaluating the network performance, capturing the propagation environment between ground BSs and both GUEs and UAVs. To do this, we adopt the newly released 3GPP 3D channel model \cite{3GPP36777}, where parameters such as path loss, shadowing, probability of line-of-sight (LoS), and small-scale fading, explicitly account for the users' height. In the remainder of the article, we provide a seminal evaluation of solutions that enable 5G-connected UAVs. The results of our extensive simulation campaigns are overviewed, explained, and finally distilled into four essential guidelines.

%% file: Section2.tex
\section*{Building the Network of Tomorrow}

\begin{figure*}[!t]
\centering
\includegraphics[width=1.9\columnwidth]{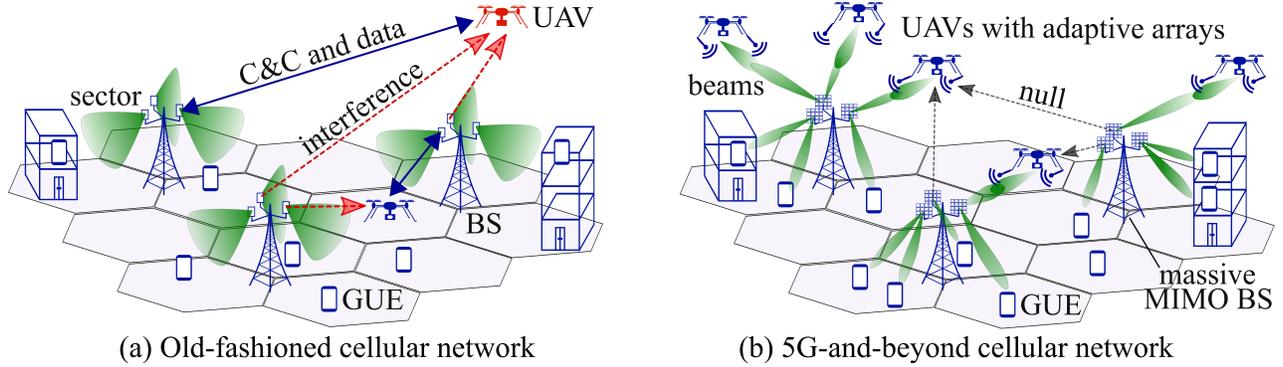}
\caption{Illustration of cellular support for UAVs through (a) an old-fashioned network and (b) a 5G-and-beyond network. In (a), each BS is equipped with a vertical antenna panel to cover a cellular sector and serve a single user on each PRB, potentially generating strong interference towards high UAVs. In (b), massive MIMO BSs serve multiple GUEs and UAVs on each PRB via digital precoding, also mitigating pilot contamination and inter-cell interference through radiation nulls, and UAVs point precise beams towards their serving BS.}
\label{fig:Network}
\end{figure*}

While we are writing this article, the cellular sites of our MNO are being upgraded from old-fashioned BSs, operating in single-user mode, to spanking new 5G massive MIMO BSs performing spatial multiplexing. Let us illustrate both modi operandi in a nutshell:\footnote{
Eager readers are referred to \cite{GerGarGal2018} for a thorough list of system parameters.}

{\bf Single-user mode (SU): }
Each BS is equipped with an $8 \times 1$ array of $\pm 45^{\circ}$ cross-polarized radiating elements, mechanically downtilted by 12$^{\circ}$. 
The radio-frequency signals at the radiating elements are combined in the analog domain, and fed to a single radio frequency (RF) chain.
Hence, the BS has one multi-element antenna which can be used to serve 
one device per time/frequency physical resource block (PRB), as illustrated in Fig.~\ref{fig:Network}(a). 

{\bf Massive MIMO mode (mMIMO): }
Each BS is still mechanically downtilted by 12$^{\circ}$, but equipped with an $8 \times 8$ array of $\pm 45^{\circ}$ cross-polarized radiating elements, as shown in Fig.~\ref{fig:Network}(b). Each element is connected to a separate RF chain, thus there are 128 single-element antennas connected to 128 RF chains.
 With this overhaul, the MNO decides to employ zero-forcing precoding and combining at each BS to spatially multiplex eight devices. To acquire channel state information (CSI), massive MIMO BSs use device-specific uplink pilot sequences, which are reused every three cells \cite{BjoHoySanBook2017}.

Common features of the SU and mMIMO paradigms are: \emph{(i)} downlink equal user power allocation, and \emph{(ii)} uplink fractional power control, where each user sets its transmit power to compensate for a fraction of the path loss incurred \cite{GerGarGal2018}.

In this article we seek to determine whether additional network or UAV enhancements are necessary to support reliable aerial links. We thus explore the following options:

{\bf UAVs with adaptive arrays (aaUAV):} 
UAVs integrate a $2 \times 2$ adaptive array comprised of omnidirectional antenna elements and a single RF chain. As illustrated in Fig.~\ref{fig:Network}(b), 
this hardware upgrade enables aerial devices to perform a precise analog beamsteering towards their serving BS.

{\bf Massive MIMO BSs with null-steering (mMIMOnulls): } 
BSs incorporate additional signal processing features that enable them to perform a twofold task. First, leveraging channel directionality, which invariably occurs in UAV-to-BS links, BSs can spatially separate non-orthogonal pilots transmitted by different aerial devices \cite{BjoHoySanBook2017}. Second, by dedicating a certain number of spatial degrees of freedom to place radiation nulls---16 in the case of our MNO---, BSs can mitigate interference on the dominant eigendirections of the inter-cell channel subspace. Such directions correspond to users in other cells that are most vulnerable to the BS's interference. Intuitively, in the presence of high UAVs, their strong LoS channels will dominate said subspace \cite{3GPP36777}. Each BS can blindly estimate the inter-cell subspace through a channel correlation estimation procedure, undertaken during silent phases \cite{GarGerGal2017}.

As it will become clear, when switching from downlink to uplink, UAVs may turn from victims of interference to offenders, at the expense of GUEs. For this reason, we also find essential to test stratagems that keep UAV-generated interference at bay:

{\bf Resource splitting (mMIMO-GUEsplit): } 
Assuming that aerial devices can be identified by the network, each BS allocates orthogonal sets of PRBs for UAVs and GUEs, making sure that the GUEs of our MNO are protected from UAV-generated interference. Let $N_{\rm UAV}$ and $N_{\rm GUE}$ be the number of active UAVs and GUEs per cell, respectively. Aiming for a fair air-time share, a fraction $N_{\rm UAV} / (N_{\rm UAV} + N_{\rm GUE})$ of PRBs is reserved to UAVs, with the remaining fraction made available solely to GUEs. Note that this may come at the expense of not fully exploiting the spatial multiplexing capabilities. The effectiveness of this approach in protecting the GUE uplink will be shown in a later section.

%% file: Section3.tex
\section*{Downlink Performance}

Putting ourselves in the MNO's shoes, we would like to understand the capabilities of reusing existing BS site locations for providing a reliable downlink C\&C channel for UAVs. Equally important---if not more---is characterizing the impact that flying devices might have on the performance of conventional terrestrial users, the latter representing the MNO's existing customers. These aspects are addressed throughout the present section.

\subsection*{UAV Downlink C\&C Channel}

\begin{figure*}[!t]
\centering
\includegraphics[width=1.9\columnwidth]{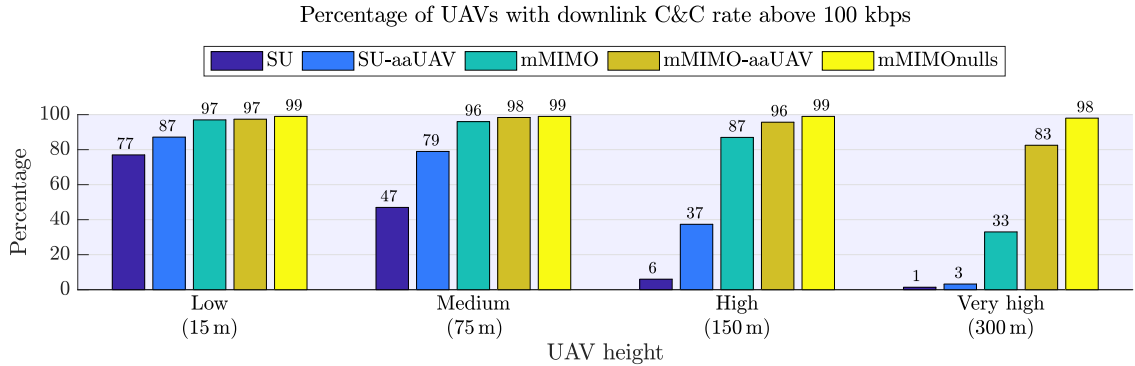}
\caption{Percentage [\%] of UAVs with downlink C\&C channel rates larger than 100~kbps as a function of their flying altitude.}
\label{fig:DL_UAV}
\end{figure*}

Fig.~\ref{fig:DL_UAV} illustrates the percentage of UAVs that achieve a downlink C\&C channel rate larger than the minimum requirement of 100~kbps \cite{3GPP36777}. This percentage is shown by considering one UAV per cell, 
and by varying the height of all UAVs (15, 75, 150, and 300~m) to exemplify the crucial role of this parameter on the ground-to-air link performance. 

Importantly, Fig.~\ref{fig:DL_UAV} demonstrates that, irrespective of the network and device capabilities, the performance of the downlink C\&C channel diminishes as UAVs increase their height from 15~m to a maximum flying altitude of 300~m. This degradation can be intuitively explained as follows:
\begin{itemize}
\item 
The MNO's BS antenna panels are downtilted, i.e., with the main lobe of the antenna pattern looking towards the center of their coverage area. Due to the strong antenna directivity, users located at low heights tend to associate to the physically closest BS. Instead, devices flying at high altitudes might associate to BSs located far away, since association is governed by the sidelobes of the BS antenna pattern perceived in the sky \cite{GerGarGal2018}.\footnote{While one could think that this reduced antenna gain is the main cause for the UAV coverage problems, this is actually not the case.} This might turn the closest BS into a strong interferer, as illustrated in Fig.~\ref{fig:Network}(a) for the red-colored UAV.
\item 
As UAVs increase their height, they also enhance the probability of being in LoS with many BSs \cite{3GPP36777,MozSaaBen2018, KhaGuvMat2018}. This entails that UAVs experience a reduced path loss with a large number of BSs simultaneously, which makes aerial devices prone to perceive interference from multiple sources and thus a degraded signal-to-interference-plus-noise ratio (SINR).
\end{itemize}

The effect of interference is particularly noticeable for SU setups (with poorer interference coordination), where the percentage of UAVs with downlink rates larger than 100~kbps goes from 77\%, for the low-altitude UAVs flying at 15~m, to a mere 1\%, when UAVs fly at 300~m. This severe degradation also occurs in spite of equipping UAVs with an adaptive array (SU-aaUAV).

Altogether, Fig.~\ref{fig:DL_UAV} demonstrates that massive MIMO could be an essential technology for providing a reliable downlink C\&C channel in highly loaded cellular networks. 
Three key factors justify the gains that massive MIMO networks provide to UAV communications:
\begin{enumerate}
\item \emph{Beamforming gains:} 
As shown in Fig.~\ref{fig:Network}(b), 
massive MIMO uses digital precoding to enhance the useful received signal strength by focusing the transmission energy on the physical UAV location.
\item \emph{Spatial multiplexing gains:} 
Massive MIMO BSs are capable of spatially multiplexing a large number of users simultaneously reusing the same time/frequency resources.
\item \emph{Air-to-ground spatial separation gains:} 
Massive MIMO BSs, dedicating spatial resources to terrestrial users, focus their energy towards the ground, and therefore are more unlikely to generate interference towards flying UAVs.
\end{enumerate}

The trends of Fig.~\ref{fig:DL_UAV} stress the need for our MNO to employ more sophisticated hardware and signal processing when serving aerial users. 
For instance, it can be observed that complementing conventional massive MIMO BS processing with explicit inter-cell interference suppression techniques (mMIMOnulls) is essential when catering for high-altitude UAVs. 
Indeed, these additional capabilities dramatically increase the percentage of UAVs that meet the 100~kbps requirement when these are flying at 300~m, from 33\% (mMIMO) to a whopping 98\% (mMIMOnulls). 
Even though each BS requires statistical channel knowledge from cell-edge devices associated to other BSs to perform inter-cell interference suppression, its acquisition is facilitated for higher UAVs. In fact, due to the strong channel directionality, the problem boils down to estimating the UAV angle of arrival.

Fig.~\ref{fig:DL_UAV} also tells us that, as far as UAV downlink performance is concerned, our MNO does not need to rely on devices with adaptive arrays (mMIMO-aaUAV), since the gain compared to mMIMO is minor at lower altitude and mMIMOnulls performs substantially better at very high altitudes.

Overall, Fig.~\ref{fig:DL_UAV} corroborates the effectiveness of massive-MIMO-based networks to serve UAVs in downlink. Massive MIMO will thus take center stage in the following. 

\subsection*{UAV-GUE Downlink Interplay}

Let us know take a complementary view, and assess the impact caused by the presence of UAVs on GUEs. This scrutiny is well founded, because there exists an interplay between GUEs and UAVs throughout multiple phases of communication:
\begin{enumerate}
\item \emph{CSI acquisition through uplink pilots:} In massive MIMO TDD networks, devices transmit uplink pilots prior to data transmission to facilitate the acquisition of CSI at the BS side. 
This knowledge can be subsequently leveraged for data transmission and reception purposes. 
During the CSI acquisition phase, devices located in different cells are generally forced to share the same pilot sequence to leave enough time for data communications, 
which leads to an imperfect CSI acquisition \cite{BjoHoySanBook2017}. 
Unless appropriately dealt with, this issue might be particularly problematic for ground users sharing the same pilot sequence as UAVs, since the latter experience LoS propagation conditions with a plurality of BSs, and might have the \emph{negative effect} of generating severe pilot contamination to GUEs.
\item \emph{Downlink data transmission through multi-user MIMO:} 
The data transmission phase also behaves differently when both terrestrial and aerial devices are served simultaneously through spatial multiplexing, 
mostly because UAVs and GUEs experience profoundly different propagation characteristics and need to share the total output power. 
These distinct features lead massive MIMO BSs to focus some of their radiated energy towards the sky, which could have the \emph{positive effect} of reducing the inter-cell interference generated towards ground devices.
\end{enumerate}

\begin{figure}[!t]
\centering
\includegraphics[width=\columnwidth]{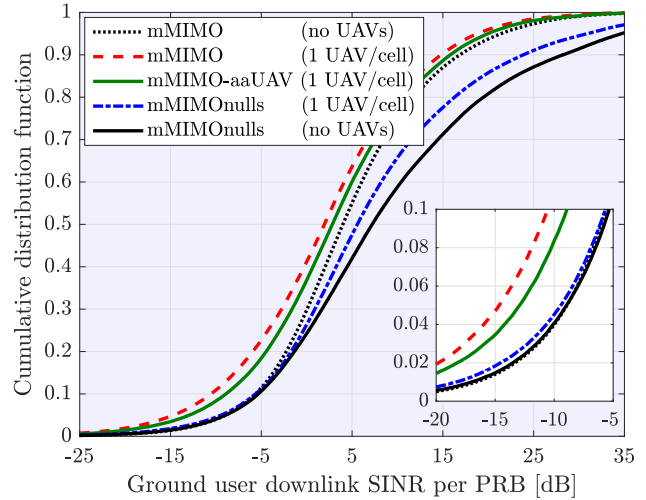}
\caption{Cumulative distribution function (CDF) of the downlink SINRs per PRB for the ground users. When present, one UAV per cell is flying at an altitude of $150$~m.}
\label{fig:DL_GUE}
\end{figure}

To illustrate the impact of the presence of UAVs on the network performance, 
Fig.~\ref{fig:DL_GUE} shows the cumulative distribution function (CDF) of the downlink SINR per PRB experienced by terrestrial users.\footnote{
A PRB occupies a bandwidth of 180~kHz and has a duration of 1~ms \cite{3GPP36777}.} Both cellular networks with and without UAVs are considered, 
with one UAV per cell flying at a height of $150$~m in the former case. 
Fig.~\ref{fig:DL_GUE} confirms that UAV-generated pilot contamination causes an overall degradation of the downlink SINRs attainable by GUEs. 
The impact is significant for cell-edge GUEs, i.e., those located in the lower tail of the CDF and shown in the inset, which lose around 5 dB when a single UAV per cell is deployed (mMIMO).

Interestingly, this performance loss is partially compensated when aerial devices are equipped with adaptive arrays (mMIMO-aaUAV), and focus their beam towards their serving BS. In fact, the increased link budget allows UAVs that perform (analog) beamforming to reduce their transmit power, in turn reducing the interference generated to other cells during the uplink pilot transmission phase. However, the MNO may not be particularly satisfied with this solution, since it still penalizes cell-edge GUEs.

On the other hand, inter-cell interference suppression capabilities (mMIMOnulls) yield a shortened gap between scenarios with and without UAVs. Fig.~\ref{fig:DL_GUE} reveals that such capabilities approximately preserve the cell-edge GUE performance even in the presence of UAVs. This remarkable result demonstrates the benefits of explicitly accounting for the presence of UAVs in the network and mitigating UAV-to-GUE pilot contamination. The remaining performance gap can be explained as follows. In the absence of UAVs, BSs point their radiation nulls towards vulnerable out-of-cell GUEs. Instead, most of the nulls target UAVs when these are present, since they are more likely to experience strong channel conditions with a large number of BSs, and are thus more vulnerable to interference.


%% file: Section4.tex
\section*{Uplink Performance}

On a par with evaluating the downlink performance of networks with UAV users, we would like to assess whether our MNO can provide high-speed uplink aerial links for video streaming purposes. An equally important concern is related to the influence that UAV transmissions have on the uplink performance of terrestrial users. These two facets of uplink transmission are treated through this section.

\subsection*{UAV Uplink C\&C Channel and Data Streaming}

\begin{figure*}[!t]
\centering
\includegraphics[width=1.9\columnwidth]{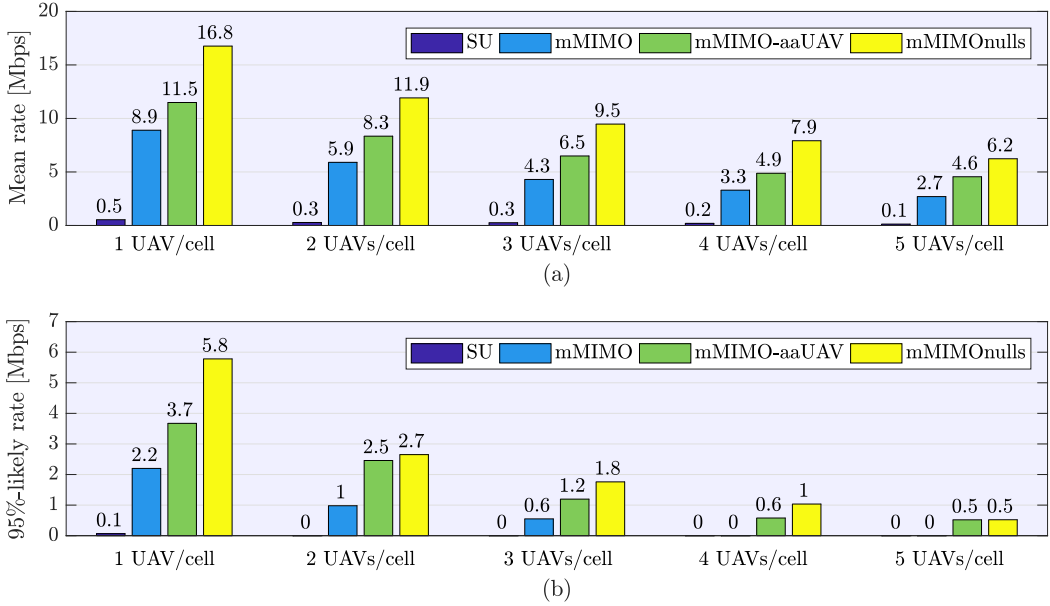}
\caption{(a) Average and (b) 95\%-likely uplink UAV rates [Mbps] for cellular networks with 1, 2, 3, 4, or 5 UAVs per cell.}
\label{fig:UL_UAV}
\end{figure*}

Fig.~\ref{fig:UL_UAV} shows both (a) the average and (b) the 95\%-likely UAV uplink data rates for a varying number of UAVs per cell. Intuitively, the average uplink data rates are indicative of the network support to real-time streaming applications, whereas the 95\%-likely rates identify the uplink C\&C channel reliability. In the considered networks, UAVs are uniformly distributed between 1.5~m, to capture their performance during the critical take-off and landing operations, and 300~m \cite{3GPP36777}.

Fig.~\ref{fig:UL_UAV} carries the consequential message that increasing the number of UAVs has a detrimental impact on their own performance. The reduction in the uplink data rates is especially significant for old-fashioned SU setups, where the minimum requirement of 100~kbps cannot be satisfied for the 5\% worst UAVs when more than one aerial user is present per cell. Therefore, once again, we shift the focal point to massive MIMO systems.

Fig.~\ref{fig:UL_UAV} confirms how 5G networks with massive MIMO BSs are capable of substantially boosting the average uplink data rates attainable by SU architectures. However, massive MIMO is not immune to severe inter-UAV interference generated both during the CSI acquisition and uplink data transmission phases. It can be observed how, with massive MIMO, the average UAV uplink rates in Fig.~\ref{fig:UL_UAV}(a) drop from 8.9~Mbps to 2.7~Mbps when the number of UAVs is increased from one to five per cell. These results hint that UAV-agnostic signal processing techniques might not be enough to guarantee the 100~kbps required by the uplink C\&C channel, when a large number of single-antenna UAVs are present in the network. 

In these challenging circumstances, the MNO could get away without null steering if all cellular-connected UAVs were equipped with adaptive arrays (mMIMO-aaUAV). But being this beyond the control of the MNO, we recommend to complement the network with inter-cell interference suppression capabilities (mMIMOnulls) to enhance the performance of the UAV uplink payload and C\&C channels. 

\subsection*{UAV-GUE Uplink Interplay}

\begin{figure}[!t]
\centering
\includegraphics[width=\columnwidth]{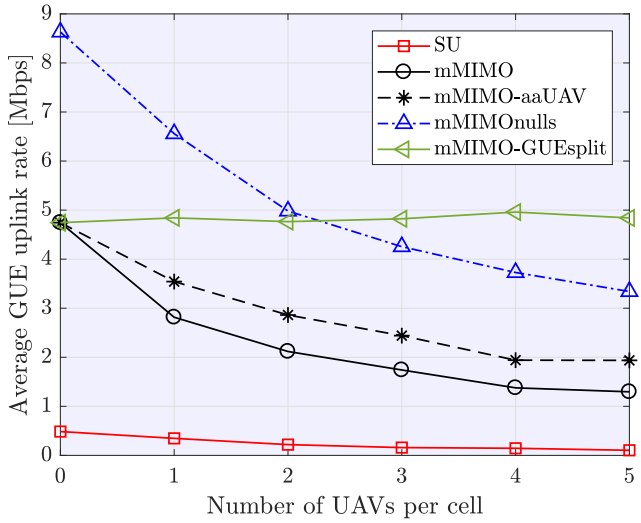}
\caption{Average uplink data rates [Mbps] for the ground users as a function of the number of UAVs per cell.}
\label{fig:UL_GUE}
\end{figure}

Fig.~\ref{fig:UL_GUE} puts a spotlight on the average uplink rates achieved by GUEs under the presence of a varying number of UAVs, from one to five per cell. Fig.~\ref{fig:UL_GUE} illustrates that the average GUE uplink rates are dramatically curtailed when more UAVs are active, except for the case when PRBs are split between GUEs and UAVs (mMIMO-GUEsplit). In this case, the performance remains roughly independent of the number of UAVs per cell.\footnote{In this example, we designed the fraction of PRBs dedicated to GUEs as a function of the number of UAVs, in order to keep the GUE rates unaltered.} Although not explicitly shown in Fig.~\ref{fig:UL_UAV} for brevity, it should be noted that this approach (mMIMO-GUEsplit) leaves UAVs with an insufficient number of PRBs, rendering it ineffective.

Fig.~\ref{fig:UL_GUE} also carries another fundamental message: GUEs may not be satisfied by the more aggressive spatial reuse offered by massive-MIMO-capable BSs, unless explicit inter-cell interference suppression mechanisms during the data reception and CSI acquisition phases are implemented (mMIMOnulls). 
Still, even this approach suffers when there are three or more UAVs per cell in a fully loaded network. 
Finally, Fig.~\ref{fig:UL_GUE} shows that equipping UAVs with adaptive arrays offers a limited performance improvement, in spite of allowing them to reduce the interference they generate thanks to both 
\emph{(i)} their reduced radiated power as per the fractional power control logic applied, 
and \emph{(ii)} the increased directionality of their transmissions. 
Altogether, Fig.~\ref{fig:UL_GUE} highlights the need to account for the presence of UAVs throughout the network design stage, 
if the performance of the existing GUEs is to be preserved.

%% file: Section5.tex
\section*{Guidelines for Realizing 5G-Connected UAVs}

\begin{table*}
\centering
\caption{Recap of network and UAV upgrades, along with their degree of effectiveness: insufficient (\xmark), partial ($\mathbf{\sim}$), or satisfactory (\cmark).}
\label{table:solutions}
\renewcommand{\arraystretch}{1.5}
\setlength\tabcolsep{5pt}
\begin{tabular}{|c|c|c|c|c|p{8cm}|}
\hline 
\multirow{2}{*}{\textbf{Upgrade}} & \multicolumn{4}{c|}{\textbf{Effectiveness}} & \multicolumn{1}{c|}{\multirow{2}{*}{\textbf{Caveats}}} \\\cline{2-5}
  & \textbf{DL UAV} & \textbf{DL GUE} & \textbf{UL UAV} & \textbf{UL GUE} &  \\\hline
\rowcolor{LightCyan}
\begin{tabular}[c]{@{}c@{}} Massive MIMO \\ [-0.75ex] (mMIMO) \end{tabular} & $\mathbf{\sim}$ & \xmark & $\mathbf{\sim}$ & $\mathbf{\sim}$ & \vspace{-0.5cm} Its efficacy fades for many high-altitude UAVs, unless complemented by more advanced channel estimation and precoding \cite{BjoHoySanBook2017}. \\ \hline
\rowcolor{Gray}
\begin{tabular}[c]{@{}c@{}} Resource splitting  \\ [-0.75ex] (mMIMO-GUEsplit) \end{tabular} & \xmark & \xmark & \xmark & \cmark & \vspace{-0.5cm} Need to identify UAVs through explicit signaling, location information, or recognizing UAV-specific channel features.  \\ \hline
\rowcolor{LightCyan}
\begin{tabular}[c]{@{}c@{}} Multi-antenna UAVs \\ [-0.75ex] (mMIMO-aaUAV) \end{tabular} & \cmark & $\mathbf{\sim}$ & \cmark & $\mathbf{\sim}$ & \vspace{-0.5cm} Increased UAV size, hardware, and complexity. Not all manufacturers may be willing to implement it. Not under direct control of the MNO. \\ \hline
\rowcolor{Gray}
\begin{tabular}[c]{@{}c@{}} Massive MIMO with inter-cell \\ [-1ex] interference suppression\\ [-0.75ex] (mMIMOnulls) \end{tabular} & \cmark & \cmark & \cmark & \cmark & \vspace{-0.65cm} Pilot contamination removal relies on enhanced processing leveraging channel directionality. Interference suppression needs channel estimation of out-of-cell UAVs, either via pilot coordination or blindly \cite{GarGerGal2017}.  \\ \hline
\end{tabular}
\end{table*}

In what follows, we sum up our findings, 
and provide a solution to the MNO's dilemma, 
as to whether its network will require substantial upgrades---and which ones---to support UAVs. We do so by drawing up four compelling guidelines.

{\bf Guideline~1: \emph{Take it or leave it}}\\
The complexity of the cellular network should be scaled up both with the number of connected UAVs and their maximum flying altitude. 
On the other hand, an MNO could decide to avoid this investment by restricting the maximum height at which it guarantees cellular service.

{\bf Guideline~2: \emph{It's all about focus}}\\
Owing to its capability of focusing multiple signals towards multiple users, 
massive MIMO is an important tool to achieve reliable UAV communications. Indeed, its adoption is critical to restrict the impact that UAV-generated interference has on legacy ground communications.

{\bf Guideline~3: \emph{No pain no gain}}\\
Table~\ref{table:solutions} serves as a digest of what the network and UAV upgrades evaluated throughout this article entail, and how they pay off. Let us recap what we have observed:

\begin{itemize}
\item The efficacy of massive MIMO (mMIMO) fades in the presence of many high-altitude UAVs. Moreover, pilot contamination may pose a severe threat that, if not properly addressed, can jeopardize the performance of existing terrestrial users. 

\item The gains of resource splitting (mMIMO-GUEsplit) are confined to the uplink of GUEs, and they may come at the cost of starving UAVs of resources. Therefore, we do not regard it as a viable solution in fully loaded networks.

\item UAV manufacturers demanding cellular service at high altitudes may need to improve the hardware characteristics of their devices, e.g., by equipping UAVs with beamforming capabilities (mMIMO-aaUAV). While this solution improves UAV performance and UAV-GUE interplay up to a certain extent, it does not fall under direct control of the MNO.

\item The best bet is for the MNO to resort to inter-cell interference suppression (mMIMOnulls) to serve both UAVs and GUEs satisfactorily. As the user load increases, additional antennas can be deployed to sharpen the interference suppression capability. Still, even this approach may not suffice to handle plenty of high UAVs, calling for new network paradigms. 
\end{itemize}

{\bf Guideline~4: \emph{The sky is the limit}}\\
In a future with a rocketing number of UAVs, operators should realize that reusing the same infrastructure for both GUEs and UAVs might not be sustainable. In this case, MNOs could design novel cellular architectures, perhaps even with dedicated resources and cellular BSs pointing towards the sky.

%% file: Section6.tex
\section*{Conclusions}
The exciting era of automation is nigh, if not underway. 
As a growing number of tasks are being delegated to machines, UAVs, 
the most mobile of them all, are the logical candidates to take over many such missions. 
The wireless industry at large is eyeing new business opportunities arising from the demand of fast, cable-less, and reliable exchange of air-to-ground information. 
In this article, we have made an effort to solve a cellular operator's dilemma as to what it will take to realize 5G-connected UAVs in fully loaded networks, 
without jeopardizing the performance of existing terrestrial users. 
Our extensive 3GPP-compliant simulations have shown massive MIMO to be a viable solution, 
if complemented with appropriate infrastructure and signal-processing upgrades by both operators and UAV manufacturers.